\documentclass[aps,prl,twocolumn,showpacs,balancelastpage,groupedaddress]{revtex4}

\usepackage{graphicx}
\usepackage{amsmath, amsthm, amssymb}
\usepackage{epsfig}
\usepackage{latexsym}
\usepackage{bm}
\usepackage{color}

\begin{document}
\title{The Enigma of Entropy} 
\author{Deepak Dhar}
\email{ddhar@theory.tifr.res.in}
\affiliation{Department of Theoretical Physics, Tata Institute of Fundamental Research, Homi Bhabha Road, Mumbai-400005, India.}
\date{\today}

\begin{abstract}
This is a light-hearted take  at the the second law of thermodynamics.
\end{abstract}
\pacs{01.30.Rr, 05.70.-a}
\maketitle

        One of the most mysterious of laws of nature is the second law of  thermodynamics. There are several equivalent formulations of this law. For our present discussion, it is enough to take the Clausius formulation that says that for an isolated system evolving in time, the entropy cannot decrease. Many of you have encountered this in your B.Sc. textbooks already. Let me explain my reasons for calling it enigmatic. I do not have any answers. I just want to say why I think that it is an interesting question.

        Let me explain myself using a parable. A foreign scientist is visiting a laboratory in China. In the day time, he discusses his work with his hosts, and in the evenings, he has to find some place to eat. Unfortunately, he knows no Chinese.

        So, he goes to a restaurant, sits down on a table, and the waiter rings in the menu. It has a very long menu. All kinds of dishes listed, about one thousand in all, and the price of each in  understandable numerals. Knowing what he is willing to spend, the man selects one of the entries in the correct price range at  random, and points to it. Eventually, the waiter brings him the  food, he eats and leaves. Next day, it is the same story. The food is satisfactory and the man has no complaints.

        But, after a few days, he discovers a new restaurant in the neighborhood. It is the same price range, and same quality of food. He tells his friends that he prefers this restaurant to the  other. The friend knows how he places his order, and wants to know what is special about the second restaurant. The man says that there the menu is bigger, ten thousand entries, and “I like to have more choices”.

        Of course, one would think that the man is quite foolish. What  difference does it make that the menu has one thousand or one  million entries, after all, in one dinner, you are only going to  eat one meal! The problem is that systems with no brains at all do  exactly this. 

 As a paranthetical remark, let me add here that  description of a system of  say gas molecules in a box  ``having  a wish to'' increase their entropy  is quite inappropriate, even though    in textbooks one often encounters such expressions as ``the electron wants to go the lowest energy state''.  The electron has no brains, and in no effective sense this could be true.  Also, if everybody would want to lower their energy, where would the extra energy go?  In my parable, this was done deliberately, as  a rhetorical device. In a more serious discussion, one should ignore this  deliberate anthropomorphic flavour  added to the story.  The  amazing fact remains  that ``having more choices'' is  such an important fundamental principle of Nature.

        Consider a known mass of a dilute gas, say hydrogen, kept in a box  at a fixed temperature. Then, given this specification of the  macroscopic system, we are able to determine other properties of the system, like the pressure, or the viscosity of the gas etc…  However, given the macroscopic parameters characterizing the  system, we cannot really tell the precise position and velocity of  each individual molecule : there are very many ``microstates''  corresponding to a fixed macrostate. According to famous equation  of Boltzmann, the entropy $S$ is equal to $k \log \Omega$, where $\Omega$ is the  number of microstates available to the system. So, the entropy is not determined by the current state (the actual microstate), but  by what others it could have been in. It seems to be not a function of what is, {\em but of what could have been!}.

        Often one explains the notion of entropy by saying that it is a measure of disorder. The system is prepared in a macrostate, but it could have been in any of the microstates of the system. When the system is observed after some time, we can only say that each microstate occurs with probability $1/\Omega$. To any such probability  distribution, in which microstate $i $occurs with probability $ p_i $,  we can define a quantity, the Shannon entropy $S_{Sh} = \sum_{i} p_i  \log p_i $. But this measures the uncertainty the observer feels, given the limited information. It seems to be a measure of  disorder in the head of the observer, not in the system.

        Continuing our story, when our hero is asked to explain what is the point of having more choices, he says, ``Well, my boss in the  lab is a very nosey person. I think he wants to know exactly I  eat, even though it is none of his business. I choose the second  restaurant, just to spite him.'' This explanation seems not really  convincing. In fact, most of the time, the boss (the external  observer) does not want to know the details of what the person  ate, and maybe the hero is not really so full of negative  emotions, like spitefulness.  One is reminded of  A. Einstein's  words : Subtle is the Lord, but malicious He is not."  For the more agnostic readers, `Lord' in the above quotation may be replaced by `Nature'.

        In fact, for systems that undergo deterministic evolution, the  microstate at later time is fully determined by the present, and  there is no need to invoke any probability ideas. In this  scenario, the hero of our story has a simpler method of making the   selection : he selects the item with the right price that comes  next in the menu to what he chose on the previous day. Here he has  no choice whatsoever. In that case, why does he still prefer the    second restaurant over the first?

        That is the mystery.

\end{document}